# Exploiting Data Parallelism in the yConvex Hypergraph Algorithm for Image Representation using GPGPUs

Saurabh Jha
VIT University, Chennai, India
saurabh.jha2010@vit.ac.in

Tejaswi Agarwal
VIT University, Chennai, India
tejaswi.agarwal2010@vit.ac.in

B. Rajesh Kanna
VIT University, Chennai, India
rajeshkanna.b@vit.ac.in

## ABSTRACT

To define and identify a region-of-interest (ROI) in a digital image, the shape descriptor of the ROI has to be described in terms of its boundary characteristics. To address the generic issues of contour tracking, the yConvex Hypergraph (yCHG) model was proposed by Kanna et al [1]. In this work, we propose a parallel approach to implement the yCHG model by exploiting massively parallel cores of NVIDIA's Compute Unified Device Architecture (CUDA). We perform our experiments on the MODIS satellite image database by NASA, and based on our analysis we observe that the performance of the serial implementation is better on smaller images, but once the threshold is achieved in terms of image resolution, the parallel implementation outperforms its sequential counterpart by 2 to 10 times (2x-10x). We also conclude that an increase in the number of hyperedges in the ROI of a given size does not impact the performance of the overall algorithm.

## Categories and Subject Descriptors

I.3 [**Computer Graphics**]: Hardware Architecture –*Graphics Processor, Parallel Processing*

## 1. INTRODUCTION

Contour tracking in digital images faces generic issues which are illustrated in [1, 2, 3]. It becomes essential to solve contour tracking challenges in multiply-connected regions and regions bounded by non-Jordan curves. To overcome these challenges, Kanna et al. [1] proposed the yCHG model which is used to track the contour deterministically. Our results with the sequential implementation of the yCHG show that:
a. The runtime of the yCHG algorithm increases linearly for images up to a resolution of 2000x2000 but a significant change in runtime is observed for images with a higher resolution.
b. The runtime remains constant for images with varying number of hyperedges.

## 2. PROPOSED METHOD

To remove the data dependencies in the existing algorithm, we divide the algorithm into two steps. The first function computes the number of cut-vertices of an image in parallel by dividing the image into a number of column vectors and each column is scheduled on a separate thread on the GPU. Each thread computes the number of cut-vertices and stores the result in an array. In the second step of the algorithm, each CUDA thread checks the number of cut-vertices in the preceding column vector of the input image. If a change is observed in the number of cut-vertices, it indicates there has been a change in the number of yConvex hyperedges for that particular column vector.

## 3. RESULTS

In order to keep constant hyperedges, we take an image of a resolution of 21000x21000 and vary the resolution. Our results are graphically plotted in Figure 1. It shows performance of the proposed parallel algorithm as compared to the existing serial implementation. Our CPU implementation consists of a 2-core Intel i5 480M having a clock speed of 2660 MHz. The GPU we used is a 16 core NVIDIA GeForce 310M having a clock speed of 1468 MHz. We understand that while newer hardware (such as cards based on the recent FERMI architecture) would undoubtedly be faster, we want to show what is possible with only limited hardware investment.

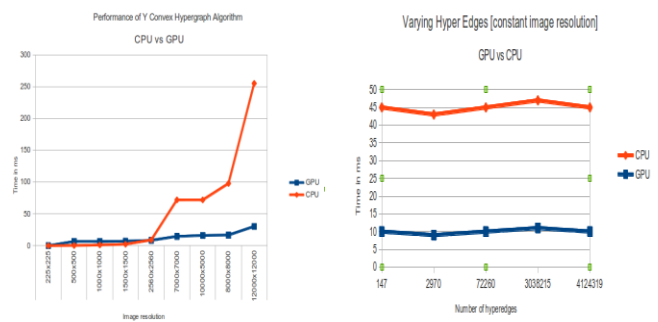

**Figure 1. Results with varying resolution and hyperedges**

To vary the number of hyperedges, we consider different images keeping the same resolution. We observe that the time taken is constant, with the number of hyperedges varying from 147 to 4124319. This was expected as our algorithm is dependent directly on the resolution of the input image irrespective of other factors. Our implementation results show that with an increase in the image resolution the parallel implementation improves the performance by 2X-10X, opening up a host of potential new applications that require real time image processing.